\documentclass[final]{elsarticle}
\usepackage{epsfig}
\usepackage{amsmath,amssymb,amsthm}
\usepackage{graphicx}% Include figure files
\usepackage{psfrag}% use latex text in ps figures
\usepackage{bm}% bold math

\begin{document}
\begin{frontmatter}
\title{Fisher-Shannon product and quantum revivals in wavepacket dynamics}

\author[1a]{T. Garc\'{\i}a}
\author[2b]{F. de los Santos}
\author[3c]{E. Romera}
\address[1a]{ Departamento de Electr\'onica y Tecnolog\'{\i}a de Computadores,
  Universidad de Granada. Fuentenueva s/n, 18071 Granada, Spain.}
\address[2b]{ Departamento de Electromagnetismo y F{\'\i}sica de la
Materia, and Instituto Carlos I de F{\'\i}sica Te\'orica y
Computacional, Universidad de Granada, Fuentenueva s/n, 18071 Granada,
Spain}
\address[3c]{ Departamento de F\'isica At\'omica, Molecular y Nuclear and
  Instituto Carlos I de F{\'\i}sica Te\'orica y Computacional, Universidad de
  Granada, Fuentenueva s/n, 18071 Granada, Spain}

%\date{\today}

\begin{abstract}

We show the usefulness of the Fisher-Shannon information product in the study of
the sequence of collapses and revivals that take place along the time evolution of quantum wavepackets. 
This fact is illustrated in  two models, the quantum bouncer and a  graphene quantum ring.

\end{abstract}
\end{frontmatter}
%\maketitle

\section{Introduction}
Sequences of collapses and revivals in the  wavepackets temporal evolution are a well
known aspect of quantum dynamics. 
This phenomenon has been  theoretically understood \cite{rob} and to date it
has been observed in striking experiments with atoms and molecules
\cite{2,nature2010}, Bose-Einstein condensates \cite{nature2002,sebby} and
recently in coherent states in a Kerr medium \cite{nature2013}. 
Moreover, quantum revivals have been studied theoretically in low-dimensional quantum
structures as graphene, graphene quantum dots and rings in perpendicular
magnetic fields \cite{romera2009,kramer2009,andrey,luo,wang,lopez, Rusin2008,Schielmann2008,Demikhovskii2012}

In this paper we show that the analysis of wavepacket quantum revivals can
be   carried out using   the  Fisher-Shannon  product $P_{\rho}$,  defined as
\cite{romera2004}:
\begin{equation}
P_{\rho}=I_{\rho}N_{\rho}
\label{fs}
\end{equation}  
with 
\begin{equation}
I_{\rho}=\int\frac{|\nabla \rho(x)|^2}{\rho(x)}dx
\end{equation}
 being the Fisher
information,
\begin{equation} 
S_{\rho}=-\int \rho(x) \ln \rho(x)dx
\end{equation}
 the Shannon entropy,  and $N_{\rho}=\exp{(2
 S_{\rho}})/(2\pi e)$ the so-called entropy power \cite{dembo}. It is known
that these 
information measures show complementary descriptions of the spreading or
concentration of the probability density, where the Fisher information gives a
local measure  of the spreading (due to the gradient in the functional form), 
whereas  the Shannon entropy provides a global one. This entropic product
has proved useful in the analysis of different physical situations, i. e.,
electronic correlation \cite{romera2004}, atomic physics \cite{nagy1,sanudo,liu},
chemical reactions \cite{esquivel},  quantum phase 
transitions \cite{nagy2}, astrophysics \cite{lovallo} or in the study of
geophysical phenomena \cite{luciano}. There is a generalization of the
Fisher-Shannon product, the so-called Fisher-R\'enyi product \cite{pla}. Note
here the existence of other important complexity measures which have also been
used in the description of a great variety of systems (see \cite{sen} and
references therein).

 We shall consider the
Fisher-Shannon  information product 
as it applies to quantum  revival phenomena. In particular, we shall show the
role  of this quantity in the dynamics of two model systems that  exhibits
sequences of quantum collapses and  revivals:  the so-called quantum
`bouncer' (that is a quantum particle bouncing against a hard surface under the
influence of gravity) and a graphene quantum ring model.

\section{Wave packet dynamic and Fisher-Shannon product}

It is well known that the temporal evolution of localized bound states $\psi$
for a time independent Hamiltonian is given in terms of the eigenvectors $u_n$ and
eigenvalues $E_n$ as
\begin{equation}
\psi(t)=\sum_{n=0} c_n u_n e^{-iE_n t/\hbar},
\label{expansion}
\end{equation}
where $c_n=\langle u_n,\psi\rangle$ are the Fourier components of the vector
$\psi$, and $n$ is the main quantum number of the system (in general
one has to consider  the set of quantum numbers corresponding to the system,
see \cite{rob}, but in this paper we will consider only systems with one quantum
number). Now,   a wavepacket is constructed with the coefficients $c_n$
 tightly centered around a large value of $n_0\gg|n-n_0|$, with  $n_0\gg 1$. The
 exponential factor in (\ref{expansion}) can then be written as a Taylor
 expansion around $n_0$ (within this approximation,  $n$ is a continuous
 variable) as   
\begin{eqnarray}
\exp\left(-i E_n t/\hbar\right)&=\exp\left[-i(E_{n_0}  + E_{n_0}^{\prime}(n-n_0) +
E_{n_0}^{\prime\prime}/2(n-n_0)^2+ \cdots)t/\hbar \right] \notag \\
=&\exp\left(-i\omega_0 t - 2\pi i(n-n_0) t / T_{\rm Cl} -2\pi i (n-n_0)^2 t /
T_{\rm R} + \cdots \right)
\end{eqnarray}
where each term in the exponential (except for the first one, which is a global
phase) defines a characteristic time scale, that is,
 $T_{\rm R}\equiv\frac{4\pi\hbar}{|E_{n_0}^{\prime\prime}|}$ and $T_{\rm Cl}\equiv\frac{2\pi
\hbar}{|E_{n_0}^{\prime}|}$ (see \cite{rob} for more details). The so called
fractional revival times can be given in terms of the quantum revival
time-scale by $t=pT_{\rm R}/q$, where $p$ and $q$ are mutually prime.

 Next, we study the wave packet dynamics by means of the so-called entropy product,  
 i.e., the product of the Fisher information  and the Shannon  entropy power, $N_{\rho}$,
 to conclude that it provides another framework for visualizing fractional revival
 phenomena. Again, we expect that the formation  of a number of minipackets of
 the original packet will correspond  to relative minima of the information
 product. Before proceeding, recall that $P_{\rho}$  satisfies the isoperimetric 
 inequality \cite{dembo} 

 \begin{equation}
 P_{\rho}=I_{\rho} N_{\rho}\geq 1.
 \label{1}
 \end{equation}
% %

 The equality is reached for Gaussian  densities.
 By combining the above inequality with the Stam uncertainty principle
 \cite{stam}  
% %

 \begin{equation}
 I_{\rho}\geq \frac{4}{\hbar^2}(\Delta p)^2,
 \label{2}
 \end{equation}
 and the power entropy inequality \cite{bia}
\begin{equation}
 N_{\rho}\leq (\Delta x)^2,
 \label{3}
 \end{equation}
% %
 leads to the usual formulation of the uncertainty principle in terms of the variance in conjugate
 spaces, $\Delta p \Delta x \geq \hbar/2$. 
 It is straightforward to show that the equality limit of these four
 inequalities  is reached for Gaussian densities.

\subsection{Quantum bouncer}

Quantum states of matter in a gravitational field have been recently realized experimentally 
with neutrons \cite{nesvizhevsky,nesvizhevsky2}. These were allowed to fall towards a horizontal 
mirror which, together with the Earth's gravitational field, provided the necessary confining 
potential well. From a theoretical point of view, this constitutes an example of the quantum variant 
of a classical particle subject to a uniform downward force, above an impermeable flat surface. 
The revival behavior of quantum bouncers has been discussed in \cite{don,gea} and an
entropy-based approach was carried out in \cite{rom1,rom2}. Here, we shall evaluate the 
goodness of the entropy-product when applied to a quantum bouncer.

The time dependent solution of the Schr\"odinger equation for the potential 
$V(z)= mgz$ if $z \geq 0$ and $V(z) =0$ otherwise reads 
\begin{equation}
\Psi(z,t) = \sum_{n=1}^\infty c_n e^{-iE_nt/\hbar} \varphi_n(z),
\label{wavefunction}
\end{equation}
where the eigenfunctions and eigenvalues are given by \cite{gea}
\begin{equation}
E'_n=z_n; \quad \varphi_n(z')= {\mathcal N}_n
{\rm Ai}(z'-z_n); \quad n=1,2,3,\ldots.
\end{equation} 
In the above equation, position and energy variables have been rescaled, $z^{\prime}=z/l_g$, $E^{\prime}=E/mgl_g$,
and are denoted by primed symbols. $l_g=\left(\hbar/2gm^2\right)^{1/3}$ is a characteristic gravitational length, 
${\rm Ai}(z)$ is the Airy function, $-z_n$ denotes its zeros, and ${\mathcal N}_n$ is the $\varphi_n(z^\prime)$ normalization 
factor. $z_n$ and ${\mathcal N}_n$ were determined numerically by using scientific subroutine libraries for
the Airy function, although accurate analytic approximations
for them can be found in \cite{gea}. In the remainder of this paper, the primes on the energy and position variables will be 
omitted and we shall assume initial conditions that correspond to Gaussian wave packets localized at a height $z_0$ above the 
surface, with a width $\sigma$ and an initial momentum $p_0=0$,
\begin{equation}
\Psi(z,0)=\left(\frac{2}{\pi \sigma^2}\right)^{1/4} e^{-(z-z_0)^2/\sigma^2}.
\end{equation}

The corresponding coefficients $c_n$ of the wave function (\ref{wavefunction}) can be obtained analytically as \cite{vallee}, 
\begin{eqnarray}
c_n= &&{\mathcal N}_n \left(2 \pi \sigma ^2\right)^{1/4} 
\exp\left[\frac{\sigma^2}{4}\left(z_0-z_n+\frac{\sigma^4}{24}\right)\right] \nonumber \\
&& \times {\rm Ai}\left(z_0-z_n+\frac{\sigma^4}{16} \right),
\end{eqnarray}
with ${\mathcal N}_n=|{\rm Ai}'(-z_n)|$. It is now a straightforward calculation to obtain for the classical period and the 
revival time $T_{\rm Cl}=2\sqrt{z_0}$ and $T_{\rm R}=4 z_0^2/\pi$, respectively \cite{gea}.
The temporal evolution of the wave packet in momentum-space is obtained numerically by the fast 
Fourier transform method.

\begin{figure}  
\includegraphics[width=0.85\textwidth,angle=0]{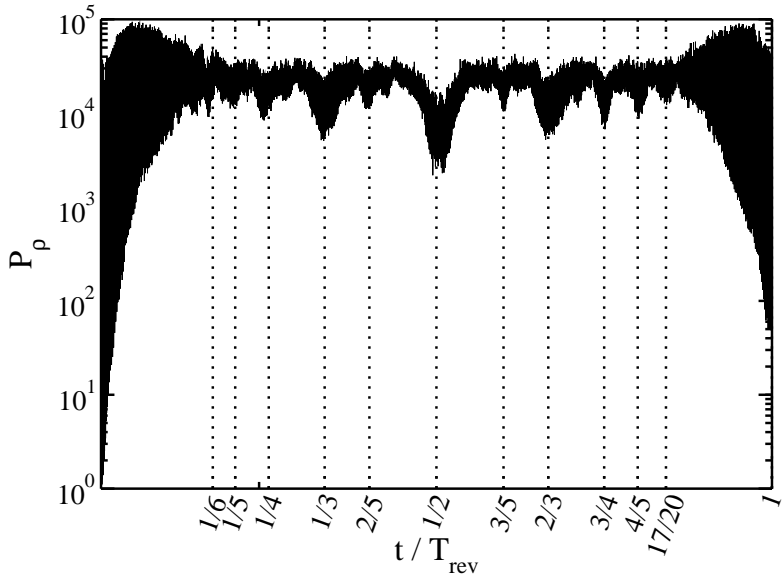} 
\caption{Time dependence of the Fisher-Shannon product for an initial 
 Gaussian wave packet with $z_0 = 100, p_0 = 0$, and $\sigma = 1$ in a quantum bouncer.
The main fractional revivals are indicated by vertical dotted lines.
} 
\label{fig1}
\end{figure}

We have computed the temporal evolution of the entropy product (\ref{fs}) for the initial conditions $z_0=100$, $p_0=0$, and 
$\sigma=1$. Figure \ref{fig1} displays $P_\rho$ and the location
of the main fractional revivals.
It can be neatly observed that the entropic product decreases and reaches a
minimum at most of the fractional revivals, where the quasiclassical behavior
and a Gaussian shape
are recovered.

Notice how the initial value of unity for the information product is approximately recovered at the full revival, 
when the Gaussian form of the wavepacket is roughly restored. 

\subsection{Graphene quantum ring}

We shall consider the behavior of the Fisher-Shannon information in another physical
situation. Let us consider a graphene quantum ring within a simplified model
\cite{4,28,29,trinidad13} which is described by the Hamiltonian
\begin{equation}
 H=v_F\vec{p}\hspace{0.05cm} \vec{\sigma} + \tau\Delta\sigma_z,
\label{Dirac}
\end{equation}
where $\tau=\pm 1$ corresponds to each of the two inequivalent corners  $K$ and $ K^{\prime}$ of the first 
Brillouin zone,  $ \vec{p}$ is the momentum measured relatively to the $K$ ($K^{\prime}$) point,
% $\vec{A}$ is the
%vector potential, 
$\Delta$ a finite mass term, $v_F\simeq 10^6$ m/s
is the Fermi velocity, and $\vec{\sigma}=(\sigma_x,\sigma_y,\sigma_z)$ where the
components are the Pauli matrices. We shall work using a geometrical approximation of a
 zero width ring with radius $R$ (used in \cite{4,28,29,trinidad13}).  The eigenstates and
 eigenfunctions are given in polar coordinates by \cite{4}
\begin{equation}
\Phi_m(R,\phi)= \left( \begin{array}{cc}
                    \phi_A(R)e^{im\phi} \\
                    i \phi_B(R)e^{i(m+1)\phi} \\
                    \end{array}
                    \right),
\label{spinor1}
\end{equation}
\begin{equation}
\phi_A(R)=1, \quad \quad  \phi_B(R)=\frac{m+\frac{1}{2}}{\epsilon+\tau \nu},
\end{equation}
\begin{equation}
 \epsilon=\pm\sqrt{(m+1)m+\nu^2+1/4},
\label{ecuacion_energia_mono}
\end{equation}
where $E_0\equiv\hbar v_F/R$,
$\nu\equiv\Delta/E_0$ and $\epsilon\equiv E/E_0$, and with
$\hbar (m+1/2)$ with $m=-j$, $-j + 1$, $\ldots j$ being the eigenvalue of the total angular momentum
$J_z=L_z + \hbar S_z$.

\begin{figure}
\centering
\includegraphics[width=0.95\columnwidth,keepaspectratio]{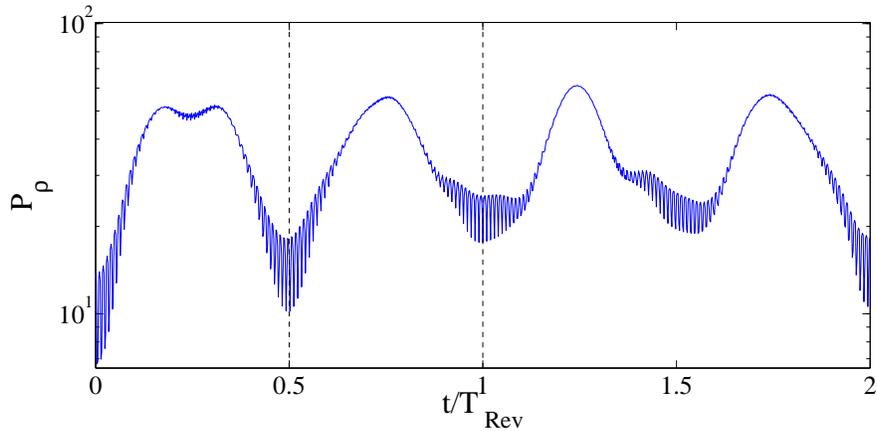}
 \caption{Time dependence of the entropic product for an initial Gaussian wavepacket 
with  $E_{m_0}\approx 200$ meV, $\sigma=13$, 
for a quantum ring with $R=50$ nm, $\Delta=50$ meV.} 
 \label{ring}
\end{figure}

Now we shall construct the initial wave packet centered around  an eigenvalue
$E_{m_0}$. In Fig. \ref{ring} the Fisher-Shannon product for a
ring of $R=50$ nm, $\Delta=50$ meV and $E_{m_0}\approx 200$ meV is depicted. We can observe a
quasiclassical evolution  (with a period of $T_{\rm R}=0.078$ ns) at early times,
which corresponds to a minimum in the Fisher-Shannon product. We can see
that at  $T_{\rm R}/2$ (and at multiples times of it), where the wave-packet recovers its quasiclassical
behavior, we have a relative minimum again.

\section{Summary}

We have 
presented an analysis of quantum wavepacket revival phenomena in the quantum bouncer and a
graphene ring, based on the information product.  We have shown that this
theoretical tool (which has proved to be useful for the analysis of different phenomena in
atomic physics, molecular reactions, solids, and even in geophysics)
appropriately describes the dynamics of wavepackets. In particular, we have
found that the revivals and fractional revivals  of wavepackets correspond to  relative  minima
in the entropic product, signaling the recovery of the quasiclassical
behavior of the wavepacket.

\section{Acknowledgments}
This work was supported by the Spanish Projects No.
MICINN FIS2009-08451, No. FQM-02725 (Junta de Andaluc\'ia), 
 and No. MICINN FIS2011-24149.

\end{document}